\documentclass{article}
\usepackage[utf8]{inputenc}
\usepackage{setspace}
\usepackage{graphicx} 
\usepackage{upgreek}
\usepackage{xcolor}
\usepackage{amsmath,amssymb,amsfonts}
\usepackage[a4paper, margin=1in]{geometry}
\usepackage{authblk}
\usepackage{CJKutf8}
\usepackage[style=apa, backend=biber]{biblatex}
\bibliography{./bibliography.bib}

\title{Correlation of Divergency: Being Different in a Similar Way or Not}
\author[1,2,3,4]{Johan F. Hoorn}
\author[1]{Johnny K. W. Ho}

\affil[1]{School of Design, The Hong Kong Polytechnic University, Hung Hom, Hong Kong SAR}
\affil[2]{Department of Computing, The Hong Kong Polytechnic University, Hung Hom, Hong Kong SAR}
\affil[3]{Research Institute for Quantum Technology (RIQT), The Hong Kong Polytechnic University, Hung Hom, Hong Kong SAR}
\affil[4]{Department of Communication Science, Vrije University, Amsterdam, Netherlands}
\date{}

\begin{document}

\maketitle
\begin{CJK*}{UTF8}{bsmi}

\begin{abstract}
Matched sets may have unlike mean scores and spreads yet share the same way in which some scores stand apart. Mean gaps and raw-score correlations do not answer this question. Correlation of divergency, $c_{\updelta},$ is put forward to fill that gap. For every score in each of two equally sized, matched sets, it finds how far that score lies from every other score in its own set. It then draws these gaps into one divergence score through a power mean. The Pearson correlation between the two lists of divergence scores is $c_{\updelta}$, which lies between $-1$ and 1. A high positive value means that the same people or things stand out in both sets, no matter which way their raw scores point. A negative value means that a score far from the rest in one set tends to be matched with a score near the rest in the other, whilst a value near zero shows little linear link. As $c_{\updelta}$ stems from within-set gaps, it is unchanged by shifts in overall level or units. It can show whether the same people feel sharply shut out in matched neighbourhood sets, whether the same bacterial strains grow oddly under two drugs, or whether the same proteins stand out in two drug-binding fingerprints. It should sit alongside means and raw-score correlations rather than take their place: $c_{\updelta}$ does not show whether an outlier is high or low, nor why it arose. The power $p$ sets the weight given to wide gaps. With $p=1$ it uses Gini distance, and with $p=2$ root-mean-square distance. Higher $p$ gives more weight to the furthest gaps but cuts Kish's effective sample size. The raw scores and the choice of $p$ must therefore be read with care.
\end{abstract}

\section{Introduction}
This paper introduces the correlation-of-divergency coefficient, $c_{\updelta}$, a custom statistical measure designed to quantify the similarity of internal divergence patterns between two groups of values. Unlike the conventional use of correlation coefficients such as Pearson or Spearman, which assess the association between paired values, $c_{\updelta}$ evaluates whether the way values differ within one group is mirrored in another. The method involves calculating, for each value, its divergence from all other values in its group, and then comparing these patterns across the two groups. The correlation-of-divergency coefficient $c_{\updelta}$ analyses the structure of the variance rather than the mean. The coefficient is normalised by an average divergence measure within each group, ensuring scale invariance.

\subsection{Different means and scales, standing out the same}
Take two equally sized, matched samples from the Hong Kong population. One comes from a low-socioeconomic status (SES) neighbourhood like Tai Kok Tsui and includes two street sleepers. The other comes from a mainly middle-class neighbourhood, the Kingswood Villas, but includes two low-SES people who have a hard time, though housed. Each person gives a score for financial anxiety or for feeling shut out.

A comparison of means will mostly show the broad gap between the groups. Pearson or Spearman correlation looks for paired scores that rise and fall together. Neither tells us whether the street sleepers stand apart from their low-SES peers in the same way that the two poorer, housed people stand apart from their Kingswood peers.

The correlation of divergency, $c_{\updelta}$, asks that question. For each person, it finds how far their score lies from every other score within their own group and merges those gaps into one divergence score. It then correlates the divergence scores across the matched samples. A high $c_{\updelta}$ means that the paired people who stand out in one group also tend to stand out in the other, even when the groups have very different means and scales. It does not equate homelessness with low pay. It shows that the within-group pattern of standing apart is alike.

\subsection{Both ostracised but in a different way}
Let us now look at two groups of the same size of older women in Hong Kong, matched by age and household size. The low-income group includes two women who gather waste cardboard for cash, often called 紙皮婆婆\ (cardboard grannies). The subdivided-flat group includes two women whose household income has fallen, but they do not live on the streets. Ask each woman how strongly she feels shut out by those around in her neighbourhood.

The cardboard gatherers may be chased from shop fronts, their equipment damaged by construction workers, looked down on, and left out of neighbourhood life. Their scores on the Ostracism Experience Scale (OES) (\cite{ostracism2025}) lie far above those of the other low-income women. The two subdivided-flat women, by contrast, may still have close ties with kin, friends, and neighbours. Their slightly higher ostracism scores remain much like those of the rest of their set.

The $c_{\updelta}$ coefficient looks at how far each woman stands from others in her own group. Here, the strongest outliers in the low-income group line up with women who barely stand out in the set for subdivided-flats. If this reverse pattern holds throughout the matched cases, $c_{\updelta}$ is negative. If it is less steady, $c_{\updelta}$ will be close to zero.

The finding would show that being cast out takes a different shape in each walk of life. The cardboard grannies stand sharply apart, while the mothers in the subdivided-flats feel less cut off.

\subsection{Different mean growth with a different set of outliers}
Suppose the Prince of Wales Hospital has seeded two Petri dishes at matched spots with the same bacterial strains taken from an asylum seeker on a hospital ward. One dish holds experimental drug A and the other holds known drug B. After a day, the lab team gauges how much each strain has grown.

Experimental drug A may slow down growth far more than drug B, so the two dishes have very different mean growth. That alone does not show whether the same strains act oddly under both drugs. The $c_{\updelta}$ coefficient may fill that gap. It works out how far each strain's growth lies from that of all the other strains in its own dish, then checks whether these within-dish gap patterns match.

A high $c_{\updelta}$ means that the same strains stand out under both drugs. If the raw results show strong growth in both dishes, those strains may bear cross-resistance and should be studied at once. A low $c_{\updelta}$ means that each drug brings out a different set of outliers, which may point to different forms of resistance and help guide drug choice.

What $c_{\updelta}$ does not do, by itself, is to show whether an outlier grows more or less. The lab must read it alongside the raw growth figures. Its use is to spot shared patterns of unusual drug response that mean growth and standard correlation may hide.

\subsection{Standing out from your own group, even when data points rise and fall in different ways}
As prescribed by the Hospital Authority, a public specialist outpatient clinic may choose from diazepam, lorazepam, buspirone, and sertraline to ease the anxiety of frontline social workers. A university lab tests each drug against the same set of brain-linked proteins and records binding strength. Suppose now that each drug has its own binding fingerprint.

Mean scores show overall binding strength. Standard use of the Pearson correlation asks whether two fingerprints rise and fall together. The analyst may instead wish to know whether the same proteins stand well apart from all the others within each drug's fingerprint, even when the readings point in unlike ways.

For each drug, $c_{\updelta}$ may work out how far each protein's reading lies from every other reading for that drug. It then compares these gap patterns two drugs at a time. Four drugs give six $c_{\updelta}$ scores. Say diazepam and lorazepam have a high $c_{\updelta}$. The same GABA-A receptor tests stand apart in both fingerprints. If the raw readings also show strong binding, this fits their shared way of working. Buspirone and sertraline might have a $c_{\updelta}$ near zero because their standout proteins do not match. A high $c_{\updelta}$ between drugs thought to work in unlike ways would give the lab grounds to check for a shared target or off-target.

In drug chemistry, $c_{\updelta}$ can help sort compounds by the shape of their binding fingerprints, indicate unwanted targets, and show whether a new molecule keeps a wanted pattern whilst shedding a harmful one. Disclaimer: $c_{\updelta}$ does not show binding strength or direction, nor which drug will ease anxiety best. Raw readings, dose-response work, safety checks, and clinical findings are still needed.

\section{The $c_\updelta$ coefficient}
Correlation of divergency, $c_\updelta$, is proposed to measure the association between patterns of divergence within a group of paired values ($x$ and $y$). It compares how each value in $x$ differs from all other values in $x$, and does the same for $y$, then assesses whether these patterns of divergence are related. The signal of divergence due to a point $i$ in the set along the $\alpha$-axis ($\alpha = x,y$) is defined using the power mean as
\begin{equation}
    D_{\alpha,i} = \left(\frac{1}{n-1}\sum_{j\ne i}^n \left|\alpha_i - \alpha_j\right|^p\right)^\frac{1}{p} \ge 0,\quad p\ne 0
\end{equation}
%
In particular,
\begin{equation}
    D_{\alpha,i} = \begin{cases}
        \displaystyle\prod_{j\ne i}^n \left|\alpha_i - \alpha_j\right|^\frac{1}{n-1},\quad p\rightarrow 0 \\
        \displaystyle\max_j|\alpha_i - \alpha_j| , \quad p\rightarrow +\infty \\ 
        \displaystyle\min_j|\alpha_i - \alpha_j| , \quad p\rightarrow -\infty
    \end{cases}
\end{equation}
No equal values are assumed for $p\le 0$. The divergence maps to the root-mean-square distance for $p=2$, and Gini's mean difference for $p=1$. For a related divergence pattern, the characteristic distance from a point with other data points would be correlated for the $x$ and $y$ dimensions. Normalising the variance of the sampling points with the respective axes, we define
\begin{equation}
    c_\updelta = \mathrm{corr}\left(D_{x},D_{y}\right) \in \left[-1,1\right].
\end{equation}
Since natural variances exist in datasets generally (distinct data points), and the variance depends on the sample size and distribution, we adapt the pmcc formalism instead of Tucker's congruence coefficient (uncentred cosine). For Tucker's congruence coefficient, the expectation of the random variable is not subtracted by mean, i.e., relative to zero, which has a particular physical significance in its own right, but $c_\updelta$ does not concern the scale of variance in the divergence space. A large $c_\updelta$ indicates that the patterns of divergence within $x$ and $y$ are similar, such that when a value in $x$ is distant from others, the corresponding value in $y$ is likewise distant (being different in the same way). Conversely, a low value suggests that the patterns of divergence are dissimilar or unrelated (being different in a different, perhaps unique, way). A negative $c_\updelta$ indicates that a point with a large deviation from the $x$ cluster tends to pair with a small deviation from the $y$ cluster across the sample, and vice versa. Like all pmcc's, $c_\updelta$ is scale-invariant because the denominator normalises the background signal of divergence with the standard deviation. $c_\updelta$ is generally not invariant to nonlinear monotone transforms. 

Instead of finding the Pearson product-moment correlation coefficient (pmcc) between $X$ and $Y$, we evaluate the pmcc of the \emph{divergences} of $X$ and $Y$. Ordinary pmcc asks whether $X$ and $Y$ follows some established relation, but $c_\updelta$ asks whether the same cases are the extreme ones in both groups, regardless of which direction `extreme' points. For example, for sampling points of a noisy square wave signal of one period $(X,Y)$, the pmcc is high but $c_\updelta$ is poor. The divergence in the $y$-axis is consistently larger because of the clustered points at either the low or high level, but $X$ is sampled in a continuous space. The different patterns of divergence would suggest that $X$ and $Y$ may differ in the marginal shape (e.g., $X$ by uniform sampling and $Y$ is threshold-based bimodal). $c_\delta$ is distinct from the original pmcc by inspecting the $x$ and $y$ variances between the data points. This gives second-order details of whether the data points reflect mapped similarity under linear transformations. 

The value of $p$ modulates the weights of the rare extreme values. A larger $p$ weights more on the extreme differences, leading to a heavier tail dependence. This increases the sensitivity to having an extreme point that is far from the cluster. The value of $c_\updelta$ is large in this case even though that extreme point aligns well with the correlation within the cluster. Then, we can detect a special point that may actually be treated differently. However, large $p$ reduces the Kish's effective sample size (alternatively, inverse participation ratio) of the $i$th data point in the dataset:
\begin{equation}
    n_{\mathrm{eff},i} = \left(n-1\right)\min\left\{\left(\frac{M_p\left(X_i - X\right)}{M_{2p}\left(X_i - X\right)}\right)^{2p}, \left(\frac{M_p\left(Y_i - Y\right)}{M_{2p}\left(Y_i - Y\right)}\right)^{2p} \right\},  
\end{equation}
where $M_q\left(\Delta\xi_i\right)$ is the power mean of $\Delta\xi_i = \xi_i - \xi$ ($\xi \ne \xi_i$) with the exponent $q$ (e.g. for $p \rightarrow \pm\infty$, $D_{\alpha,i}$ reduces to the distance between the extreme point and point $i$, as if there is only one data point), and thus the effective sample size of the entire sample:
\begin{equation}
    n_{\mathrm{eff}} = \bar{n}_{\mathrm{eff},i} = \frac{1}{n} \sum_{i=1}^n n_{\mathrm{eff},i}.  
\end{equation}
Therefore, a careful choice should be made to balance the sensitivity and the effective sample size.


\section{Discussion}
The bespoke correlation-of-divergency coefficient, $c_{\updelta}$, investigates the structure of variance among groups of data, asking whether the same matched cases stand apart within their own sets or not. For each case, it draws the gaps from all its ingroup-members into one divergence score through a power mean. It then correlates the two lists of divergence scores. This can bring to light a shared shape of standing apart even when the sets have unlike overall levels, spreads, units, or raw-score trends. The $c_{\updelta}$ coefficient should therefore work alongside means, Pearson or Spearman correlation, and plots rather than take their place.

Adding the same fixed amount to every score in either set leaves $c_{\updelta}$ unchanged. So does rescaling either set by any non-zero amount, as long as the divergence scores still have some spread. Swapping $x$ and $y$ also gives the same answer. These traits make $c_{\updelta}$ useful when matched readings come from tools or settings with different baselines and units. No bell-shaped spread is needed merely to work out the coefficient, nor must the raw $x$ and $y$ scores follow a straight-line link.

The coefficient also has a fairly plain reading. A high positive $c_{\updelta}$ means that cases far from their peers in one set tend to be far from their peers in the other. A negative value means that cases which stand out in one set tend to lie near the main body of the other. A value near zero shows little straight-line link between the two lists of divergence scores. The power $p$ gives further control. With $p=1$, divergence rests on mean absolute, or Gini, distance; with $p=2$, it rests on root-mean-square distance. Larger values give more weight to the widest gaps. This lets the analyst tune $c_{\updelta}$ towards broad differences or the furthest cases.

\subsection{Limitations}
The strengths discussed previously arise because $c_{\updelta}$ narrows the question, but that narrowing also throws away detail. Each case has $n-1$ pairwise gaps, yet all of them are folded into one number. Two sets may therefore have much the same divergence scores whilst their full webs of pairwise gaps are quite unlike. A high $c_{\updelta}$ shows shared overall remoteness among matched cases. It does not show that every gap matches, that both sets arose through the same data-making process, or that one brought about the other. Since $c_{\updelta}$ is a Pearson correlation of divergence scores, a value near zero may also hide a curved or other non-linear link between them.

Direction is lost as well. A score far above its peers and one far below them may have much the same divergence score. By itself, $c_{\updelta}$ cannot tell whether a standout case is high or low, welcome or harmful, growing or shrinking. Raw scores and plots are needed for that. The measure is also global rather than local (except for $p \rightarrow -\infty$, resembling the nearest-neighbour distance) because each case is compared with the whole set. Small clusters or close-range patterns may be missed when their members have much the same overall distance from all other cases.

The choice of $p$ is both a strength and a risk. A large $p$ may let one far-off score shape many divergence scores, since that score enters the gaps for every other case. The effective sample size may then fall sharply. At $p \leq 0$, ties and zero gaps need careful handling, while $p=0$ and $p=\pm\infty$ require clear limiting rules. Trying many values of $p$ and reporting only the strongest $c_{\updelta}$ would overstate the finding. There is also no sound cut-off at which $c_{\updelta}$ becomes "high". Its likely range and steadiness depend on the sample size, the score spread, ties, $p$, and the way the data were gathered.

Additionally, $c_{\updelta}$ cannot be worked out when either list of divergence scores has no spread. This always occurs with only two cases and may also arise in tied or highly even data. Small samples are likely to yield an unsteady coefficient. Moreover, each raw score feeds into nearly every divergence score, so the resulting case-level scores are not independent. Ordinary Pearson $p$-values and confidence bounds should not be carried over without proof that they hold. Sound matching is vital too. Missing cases, wrong pairings, repeated records and clustered data may all skew the result. A na\"{i}ve calculation also takes work that grows with $n^2$, which may become burdensome for very large sets.

\subsection{Application areas}
The same question could be useful in many fields beyond the settings set out above. At river gauges, $c_{\updelta}$ could show whether the same sites stand out during two floods or in measured and forecast water levels, despite a large gap in overall flood strength. On farms, it could test whether the same plots stand apart in wet and dry years. In factory upkeep, readings from the same machines under light and heavy loads could show whether unusual vibration, heat, or power use falls on the same machines. Such findings might help choose which machines or sites should be checked first, though the raw readings would still show the kind of fault.

In schools, $c_{\updelta}$ could ask whether the same pupils stand apart in two subjects or school years, without treating a high mark as the same as a low one. In sport, it could track whether the same players stand out across drills, seasons, or playing grounds. Network researchers might use it to find whether the same devices or nodes are unusually active in two time-windows. In word studies, matched words from two bodies of text could be checked for shared patterns of standing apart in use counts. One could likewise ask whether the same stars or other sources stand out in two wavebands. In each case, $c_{\updelta}$ would mark shared unusualness, not explain its cause.

\subsection{Future work}
Further work should first show how $c_{\updelta}$ behaves from sample to sample. Where case labels may fairly be swapped, randomisation tests could break the $x-y$ matching whilst leaving each set unchanged. Bootstrap and leave-one-out schemes need more care because removing or drawing one case afresh changes every divergence score. Trials with made-up data should map bias, spread, false-alarm rates and power under small samples, ties, skew, heavy tails and measurement error. $c_{\updelta}$ should also be set against Pearson and Spearman correlation, distance correlation, Mantel tests, and simple overlap counts for flagged outliers. This would show when it adds worthwhile knowledge.

Clear guidance on $p$ is also needed. One option is to report a curve of $c_{\updelta}$ across a small, pre-set range of $p$ values rather than choose the most striking result after looking at the data. The effective sample size should be given for each set. Work is also needed on rank-based and outlier-resistant forms, a signed companion measure, weighted cases, missing scores, and records with many features. A many-feature form could replace absolute gaps with a well-chosen distance, although high-dimensional distance and scaling would then bring new problems.

Lastly, case-level displays would make $c_{\updelta}$ easier to check. An influence score could show which matched cases drive a high or negative result. Faster ways of working it out are also within reach: $p=1$ can draw on sorted scores and running sums, while $p=2$ can use sums and sums of squares. Open software should report $n$, $p$, effective sample size, warnings for ties or zero spread, and suitable plots. Used with these safeguards, $c_{\updelta}$ can offer a clear view of matched standing apart without being asked to support claims that it cannot bear.
\end{CJK*}

\printbibliography

\end{document}